\documentclass[conference]{IEEEtran}

\IEEEoverridecommandlockouts                              

\usepackage{cite}
\usepackage{amsmath,amssymb,amsfonts}
\usepackage{algorithm}
\usepackage{algpseudocode}
\usepackage{graphicx}
\usepackage{textcomp}
\usepackage{hyperref}
\usepackage{booktabs}
\usepackage[table,xcdraw]{xcolor}
\usepackage{caption}
\usepackage{subcaption}
\usepackage{float}
\usepackage{graphics}
\usepackage{multirow}
\usepackage{blindtext}
\usepackage{comment}
\def\BibTeX{{\rm B\kern-.05em{\sc i\kern-.025em b}\kern-.08em
    T\kern-.1667em\lower.7ex\hbox{E}\kern-.125emX}}

\title{\LARGE \bf MapViT: A Two-Stage ViT-Based Framework for Real-Time\\ Radio Quality Map Prediction in Dynamic Environments}

\author{Cyril Shih-Huan Hsu$^{1}$, Xi Li$^{2}$, Lanfranco Zanzi$^{2}$, Zhiheng Yang$^{1}$, Chrysa Papagianni$^{1}$, Xavier~Costa-P\'erez$^{2,3}$
\thanks{$^{1}$ Cyril Shih-Huan Hsu, Zhiheng Yang and Chrysa Papagianni are with Informatics Institute, University of Amsterdam, Amsterdam, Netherlands.{\tt \{s.h.hsu, z.yang, c.papagianni\}@uva.nl}}%
\thanks{$^{2}$Xi Li and Lanfranco Zanzi are with NEC Laboratories Europe, 69115 Heidelberg, Germany.{\tt \{Xi.Li, Lanfranco.Zanzi\}@neclab.eu}}%
\thanks{$^{3}$Xavier Costa-P\'erez is with i2CAT Foundation, NEC Laboratories Europe, and the Catalan Institution for Research and Advanced Studies (ICREA), 08010 Barcelona, Spain. {\tt xavier.costa@neclab.eu}}%
}

\begin{document}

\maketitle
\thispagestyle{empty}
\pagestyle{empty}

\begin{abstract}
Recent advancements in mobile and wireless networks are unlocking the full potential of robotic autonomy, enabling robots to take advantage of ultra-low latency, high data throughput, and ubiquitous connectivity. However, for robots to navigate and operate seamlessly, efficiently and reliably, they must have an accurate understanding of both their surrounding environment and the quality of radio signals.
Achieving this in highly dynamic and ever-changing environments remains a challenging and largely unsolved problem.
In this paper, we introduce \emph{MapViT}, a two-stage Vision Transformer (ViT)-based framework inspired by the success of pre-train and fine-tune paradigm for Large Language Models (LLMs).
MapViT is designed to predict both environmental changes and expected radio signal quality.
We evaluate the framework using a set of representative Machine Learning (ML) models, analyzing their respective strengths and limitations across different scenarios.
Experimental results demonstrate 
that the proposed two-stage pipeline enables real-time prediction, with the ViT-based implementation achieving a strong balance between accuracy and computational efficiency.
This makes MapViT a promising solution for energy- and resource-constrained platforms such as mobile robots. Moreover, the geometry foundation model derived from the self-supervised pre-training stage improves data efficiency and transferability, enabling effective downstream predictions even with limited labeled data.
Overall, this work lays the foundation for next-generation digital twin ecosystems, and it paves the way for a new class of ML foundation models driving multi-modal intelligence in future 6G-enabled systems.

\end{abstract}


\section{Introduction}
\label{sec:intro}

Recent advances in mobile communication systems such as 5G and the new developments towards 6G pave the way for ubiquitous wireless connectivity in the context of robotic systems, allowing robots to leverage ultra-low latency and high data rates to enhance their autonomy. 
In scenarios involving multi-robot systems or human-robot collaboration, reliable and efficient wireless communication is essential for coordination and computational offloading to edge infrastructure. 
However, mobile robots operate in wireless environments subject to dynamic fluctuations due to interference, reflection, and refraction amplified by the mobility of surrounding objects, posing a major challenge for reliable radio connectivity. 
Traditional approaches of Radio Quality Map (RQMap) estimation often fail to cope with these dynamic changes, leading to reduced performance and reliability. 
The ability of a robotic system to comprehend and adapt to its radio environment is a fundamental requirement for efficient operation, especially in complex, highly dynamic, and varying physical environments~\cite{GROSHEV2023166}.
Traditional RQMap estimation methods struggle to adapt in real time, especially as robots can only collect radio data along their trajectories, relying on expensive onboard sensors used to build environmental knowledge. This makes generating global RQMaps both time-consuming and inefficient, particularly in environments where radio quality fluctuates due to moving objects and changing layouts.

To address this challenge, an efficient model is needed, capable of learning the radio quality of dynamic indoor or outdoor environments, enabling the generation of real-time RQMaps of the wireless propagation environment. This would allow mobile robots to make more informed decisions and adapt their autonomous operations in response to environmental changes. The accuracy and efficiency of dynamic RQMaps estimation across space and time can significantly influence robotic decision-making processes, such as navigation planning or offloading. For example, by selecting optimal radio link conditions for computational offloading, mobile robots can conserve energy that would otherwise be spent transmitting data under poor channel conditions. To this end, we propose \textbf{MapViT}, a two-stage framework that learns the spatiotemporal dynamics of the environment and predicts the expected radio quality over space and time, thereby enhancing the autonomy and reliability of mobile robotic systems. The main contributions are summarized as follows:
\begin{itemize}
\item We propose MapViT, a novel two-stage Machine Learning (ML)-based framework for (\emph{i}) prediction of environmental changes (\emph{ii}) estimation of resultant radio signal quality. 
The ViT-based backbone is systematically compared against benchmark ML architectures to quantify the accuracy–efficiency trade-offs on different platforms.
\item We develop a Geometry Foundation Model (GFM) via self-supervised learning on depth map sequences, allowing the model to learn generalizable geometric priors. Experimental results demonstrate that the use of GFM facilitates data-efficient transfer to multiple geometry-derived modalities besides radio-quality mapping.
\end{itemize}


\begin{figure}[h]
    \centering
\includegraphics[width=1\columnwidth, clip, trim = 0cm 1cm 0cm 0cm]{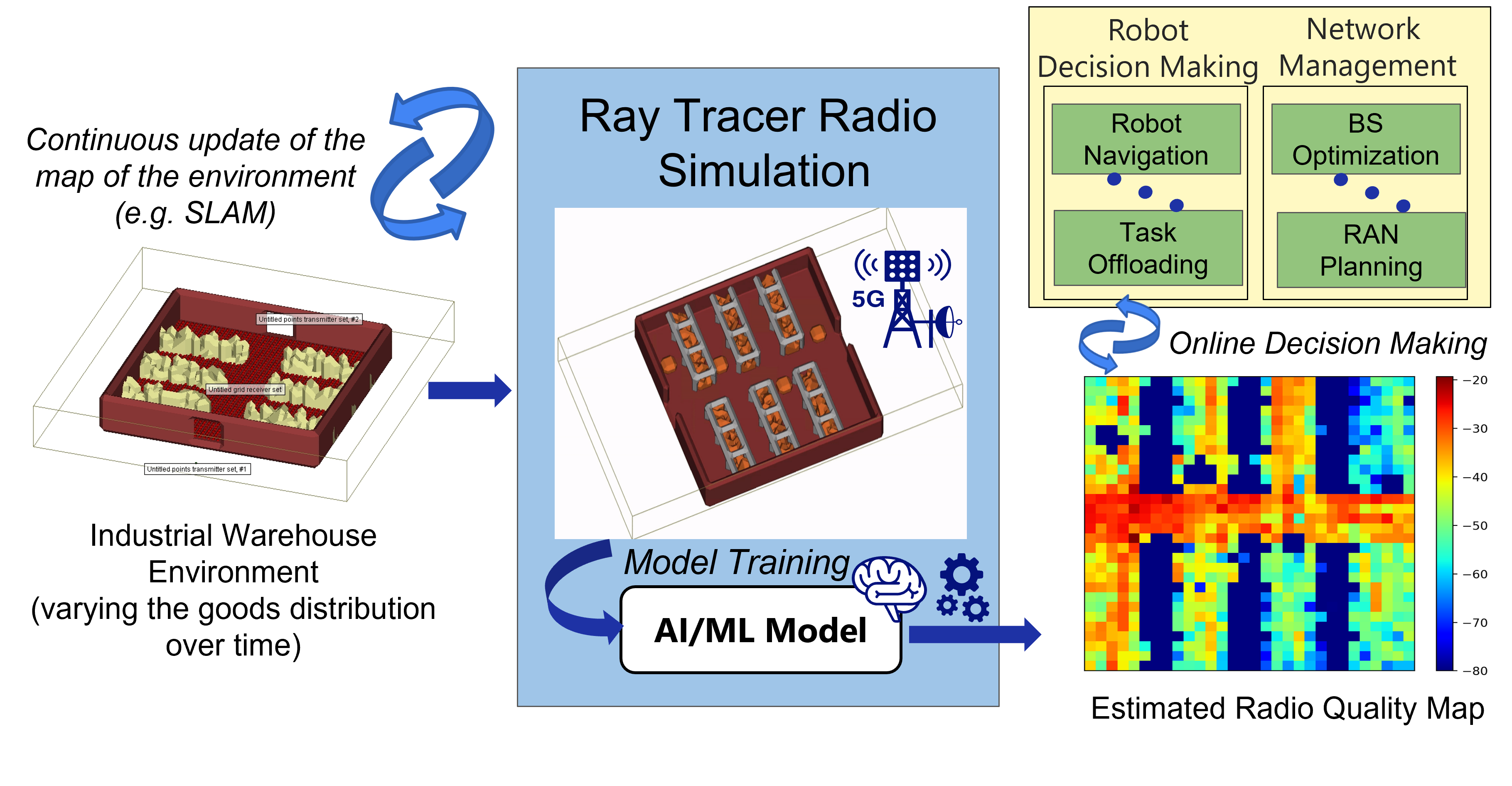}
    \caption{\small Architecture Framework.}
    \label{fig:architecture}
\end{figure}
The remainder of the paper is structured as follows. Sec.~\ref{sec:related} summarizes the related works in the field. Sec.~\ref{sec:framework} presents the proposed framework, detailing both training and deployment pipelines for prediction of environmental changes and expected RQMaps.
Sec.~\ref{sec:perf_eval} evaluates MapViT in terms of its gains, trade-offs, and data-efficiency.
Finally, Sec.~\ref{sec:conclusion} concludes this paper with insights and future research directions.

\section{Related Work}
\label{sec:related}
 
Traditional methods for radio quality prediction rely on active sensors sensing the environment~\cite{im2014radio}. 
In the context of mobile robots, the authors of~\cite{hsieh2004constructing} consider a scenario involving multiple robots collaboratively measuring signal strength at different locations. This approach assumes prior knowledge of the environmental structure, making it unsuitable for unknown environments. Furthermore, it only relies on collected data without predictive modeling capabilities. 
%
More recently, several works investigated the adoption of ML approaches to overcome hardware cost, deployment complexity, and adaptability to dynamic settings.
\cite{fink2010online} considers Gaussian Process (GP) to provide the maximum likelihood estimate of the source location and predict the received signal strength with confidence bounds in regions of the environment that have not been explored. GP models require extensive prior knowledge or environmental modeling, and their computational complexity grows significantly with increasing data volume, restricting their real-time applicability.
More recently, deep learning methods have shown potential in radio environment modeling. 
For instance, RadioUnet has proven the effectiveness of CNNs in predicting radio quality maps \cite{levie2021radiounet}. As a drawback, CNNs struggle with capturing long-range dependencies, which is crucial for global contextual understanding in dynamic environments where signal conditions change over time.

To address dynamic environmental changes, researchers have explored alternative learning paradigms.
OnRMap~\cite{dos2023onrmap} adopts a self-learning paradigm that dynamically updates predictions based on real-time observations and environmental changes.
However, OnRMap relies on frequent real-time measurements, requiring active and frequent signal acquisition to maintain accuracy, resulting in significant computational overhead.
To address these challenges, we introduce MapViT, a two-stage ViT-based framework that is capable of anticipating environmental changes as well as enhancing computational efficiency. Compared to traditional ray-tracing methods, it delivers faster performance and reduces deployment costs, without compromising prediction accuracy.

\section{MapViT Framework and Pipelines}
\label{sec:framework}

\subsection{Architectural Framework}
\label{sec:arch}
To facilitate the integration of RQMaps estimation with robotic tasks, we have devised the architectural framework illustrated in Fig.~\ref{fig:architecture}.
Dynamic environments present fundamental challenges for robot control and reliable radio quality estimation.
For example, consider an industrial warehouse characterized by long hallways and high shelves storing goods, where indoor radio connectivity is provided by a wireless antenna mounted on the ceiling (see Fig.~\ref{fig:architecture}). Within this environment, the distribution of goods on the shelves changes over time according to order demands, typically optimized by algorithms designed to maximize efficiency and minimize retrieval times. As a result of these shifts and rearrangements of goods, the signal strength and propagation patterns can fluctuate, making it difficult to maintain accurate radio quality estimates and plan robotic tasks involving computational offloading~\cite{ICRA24_I2CAT}.
A straightforward approach would involve continuously and periodically re-assessing the environment in both the physical and radio domains. The 3D map of the physical environment can be continuously collected by the robots, for instance, using Simultaneous Localization and Mapping (SLAM), as shown in the left of Fig.~\ref{fig:architecture}. This environmental map can then be fed into a radio simulator, such as a \emph{Ray Tracer}\footnote{Wireless InSite, 3D Wireless Prediction Software, https://www.remcom.com/wireless-insite-propagation-software}
, which provides 3D ray-tracing capabilities to estimate radio propagation and wave interactions with physical objects for 5G, WiFi, or other wireless networks, as illustrated in the middle of Fig.~\ref{fig:architecture}.

Nevertheless, such an approach entails substantial computational overhead, to continuously analyze the changing layout of the warehouse and optimize the radio configuration accordingly (see the right side of Fig.~\ref{fig:architecture}). 
Estimating radio propagation involves tracing multiple signal paths and computing their interactions with heterogeneous surfaces, considering physical reflections, refractions as well as other radio phenomena.
The computational complexity of these operations increases exponentially with the scene size, environmental complexity, and target resolution. In addition, uploading 3D LiDAR scans for SLAM processing requires very high transmission bandwidth, which may not always be available, particularly in areas with limited or poor wireless coverage~\cite{Romero23}\cite{5GERA_IROS}.


Therefore, in this paper, we propose a ML-based solution to alleviate the computational burden of the \emph{Ray Tracer} while maintaining high-fidelity radio quality estimation by leveraging recent advancements in deep learning. The proposed approach enables the prediction of both environmental changes and RQMaps, which can be used as inputs for real-time and autonomous decision-making in robot navigation and operations, as well as for optimizing network configurations.



\begin{figure}
    \centering
\includegraphics[width=1\columnwidth]{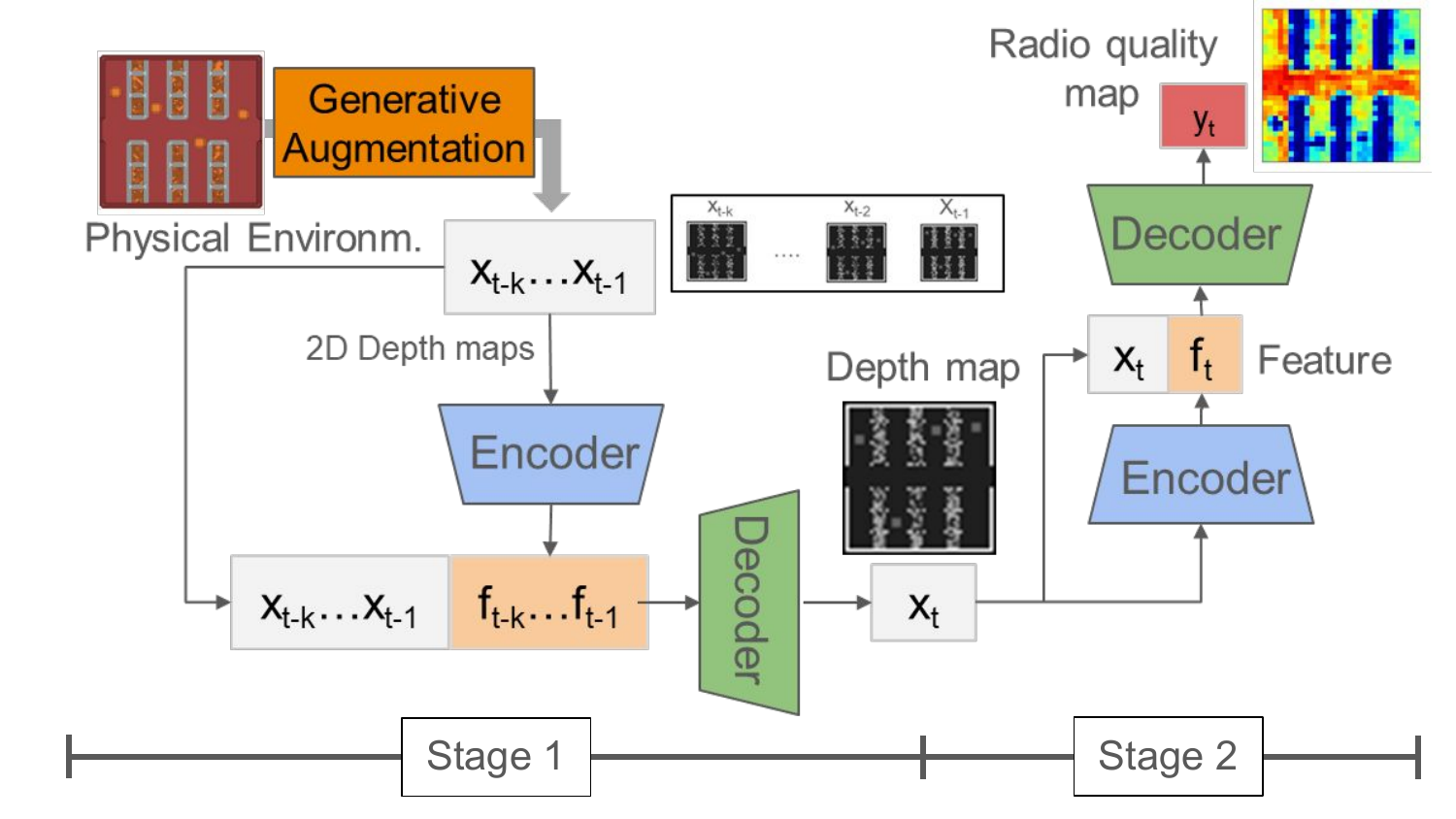}
    \caption{\small The deployment pipeline of MapViT. In Stage 1 (left), a
    sequence of previous depth maps over a given time window is processed to predict the next depth map. In Stage 2 (right), the predicted depth map is used to generate the corresponding RQMap.}
    \label{fig:pipeline}
\end{figure}

\subsection{Deployment Pipeline of AI/ML Model}
\label{sec:pipeline}
ViTs have recently emerged as a powerful alternative to CNNs in image processing, leveraging transformer architectures and self-attention to model global dependencies~\cite{dosovitskiy2021an, raghu2021do}. Their strength in capturing long-range relationships makes them well-suited for modeling RQMaps in dynamic environments, where radio propagation is affected by complex, time-varying spatial factors. By learning these dependencies, ViTs can improve the accuracy and informativeness of quality map predictions, supporting better decision-making for wireless-connected robots. 
As shown in Fig.\ref{fig:pipeline}, which provides a detailed view of the \emph{AI/ML Model} component in Fig.~\ref{fig:architecture}, the deployment pipeline consists of two stages: ($1$) predicting environmental changes at time $t$, and ($2$) estimating the expected RQMaps. Stage 1 leverages SLAM data to capture the 3D layout and goods distribution in the warehouse.
The resulting point clouds are then projected onto a 2D x-y plane to generate \textbf{Depth Maps}\cite{depth0,depth1}, where pixel intensity reflects height (z-axis), normalized to $[0,1]$ to facilitate model convergence. This transformation compresses the spatial data into a lightweight 2D representation, while preserving essential geometric features for modeling radio propagation.
\begin{algorithm}[h]
\caption{Two-stage training scheme}
\label{alg:2stage}
\begin{algorithmic}
\small
\State \textbf{Input:}
    \State  \ \ $\mathcal{M^{\prime}} = \{x^{\prime}_1, x^{\prime}_2, \dots, x^{\prime}_m\}$ \textit{// Unlabeled maps}
    \State \ \ $\mathcal{M} = \{(x_1, y_1), (x_2, y_2), \dots, (x_n, y_n)\}$ \textit{// Labeled maps}
\State \textbf{Output:}
    \State \ \ $E_d$, $D_d$ \textit{// Depth map encoder and decoder}
    \State \ \ $E_r$, $D_r$ \textit{// Radio quality map encoder and decoder}
\State \textit{// Stage 1: Self-supervised Pre-training}
    \State \textbf{Initialize} $E_d$, $D_d$
    \State \textbf{for each $epoch$ do:}
        \State \ \ \ \ \textbf{for $x_t \in \mathcal{M^{\prime}}$ do:}
            \State  \ \ \ \ \ \ \ \  $\mathbf{x}_t = [x_{t-k}, \dots, x_{t-1}]$ \textit{// Define input window}
            \State  \ \ \ \ \ \ \ \ $\mathbf{z}_t \gets E_d(\mathbf{x}_t)$ \textit{// Obtain latent codes}
            \State  \ \ \ \ \ \ \ \ $\hat{x}_t \gets D_d(\mathbf{z}_t)$ \textit{// Obtain predicted map}
            \State  \ \ \ \ \ \ \ \ $\mathcal{L}_d=\lVert\hat{x}_t-x_t\rVert^2$ \textit{// Min. loss and update $E_d$,$D_d$ }
        \State \ \ \ \ \textbf{end for}
    \State \textbf{end for}
\State \textit{// Stage 2: Supervised Fine-tuning}
    \State \textbf{Initialize} $E_r \gets E_d$, $D_r \gets D_d$
    \State \textbf{for each $epoch$ do:}
        \State \ \ \ \ \textbf{for $x_t, y_t \in \mathcal{M}$ do:}
            \State  \ \ \ \ \ \ \ \ $z_t \gets E_r(x_t)$ \textit{// Obtain latent code }
            \State  \ \ \ \ \ \ \ \ $\hat{y}_t \gets D_r(z_t)$ \textit{// Obtain predicted map}
            \State  \ \ \ \ \ \ \ \
            $\mathcal{L}_r=\lVert\hat{y}_t-y_t\rVert^2$  \textit{// Min. loss and update $E_r$, $D_r$}
        \State \ \ \ \ \textbf{end for}
    \State \textbf{end for}
\State \textbf{return} $E_d, D_d, E_r, D_r$
\end{algorithmic}
\end{algorithm}
The obtained depth map is used as input for the ViT model. The presence of mobile obstacles on the ground (e.g., humans, machinery, AGVs, etc.) and the temporal variations in goods distribution, both contributing to environmental dynamics, are captured from the depth maps of the past time slots ($t-k$, ...., $t-1$) within a predefined time window. By processing these consecutive snapshots through an aggregation decoder, the model can autoregressively predict the distribution of obstacles and goods for the next time slot $t$.

The architecture adopts an encoder-decoder design with skip connections to preserve critical spatial details. The encoder compresses the input into a compact latent representation, enabling the decoder to reconstruct the target output by leveraging learned input-output relationships. A ViT serves as the encoder, capturing global contextual features from the entire spatial domain of the input depth maps and encoding them into informative embeddings.
On the decoding side, convolutional layers map these embeddings back to the spatial domain, such as depth maps or RQMaps, depending on the stage of the process. 
By jointly processing the features extracted by the encoder and the high-resolution spatial information passed via the skip-connections, the decoder predicts the future depth map and its corresponding RQMap as the output.
To further enrich the training dataset and alleviate potential overfitting, we introduce the use of \textbf{Generative Augmentation} module to inject dynamicity into a single data instance. This module operates by intelligently manipulating the acquired geometrical data and creating diverse yet realistic variants of the warehouse environment. Each variant is then evaluated in the\textit{ Ray Tracer} to compute the corresponding RQMap. Thus, the model can be trained using the expanded dataset, comprising the original and augmented environment geometries paired with their respective RQMaps. This pipeline effectively addresses the limitations of sensor-only data by incorporating synthetic variations, improving robustness, and generalization in unseen scenarios. The outcome of the Stage 2 model is the expected RQMaps for the next time slot $t$, allowing robots to respond to environmental changes. Notably, the two-stage process is applicable to both online and offline scenarios.
\subsection{Training Pipeline}
\label{sec:training}
To train the two-stage MapViT pipeline shown in Fig.~\ref{fig:pipeline}, we combine self-supervised pre-training with supervised fine-tuning, as outlined in Algorithm~\ref{alg:2stage}. Inspired by recent advances in LLMs, the first stage involves \emph{self-supervised pre-training} on a large set of unlabeled depth maps $\mathcal{M}^\prime = {x^{\prime}_1, x^{\prime}_2, \dots, x^{\prime}_m}$, where each $x^\prime_t$ represents a depth map at time $t$. This stage trains the Stage 1 model to capture spatial relationships and temporal dynamics of the environment~\cite{vrep}.
However, since pre-training lacks direct exposure to radio signal quality, we follow with a \emph{supervised fine-tuning} stage using a smaller labelled dataset $\mathcal{M} = {(x_1, y_1), (x_2, y_2), \dots, (x_n, y_n)}$, where $x_t$ is a depth map and $y_t$ its corresponding RQMap generated via ray tracing. Fine-tuning Stage 1 on this dataset enables Stage 2 to learn the mapping between geometry and radio propagation characteristics.
This design enables dynamic RQMaps generation without needing labels for every possible future scenario. It decouples training, reducing the need for extensive labelled data while maintaining high prediction accuracy.

Overall, the proposed two-stage scheme offers the following advantages: (\emph{i}) reducing labelling effort by leveraging self-supervised pre-training, limiting the need for large amounts of labelled RQMap data;
(\emph{ii}) improved training efficiency and lower computational cost by decoupling the learning tasks; and
(\emph{iii}) creating reusable, task-adaptable foundation models that can be fine-tuned for downstream applications using small, specific datasets.
An investigation of two-stage training against a single-stage baseline is presented in Sec.~\ref{subsec:perf_data}.
%

\section{Performance Evaluation}
\label{sec:perf_eval}

We consider the indoor scenario depicted in Fig.~\ref{fig:architecture} representing an industrial warehouse of $50$m$\times50$m$\times8$m consisting of several shelves and industrial goods. The shelves are equally spaced within the area, while the distribution of goods over the shelves at each location changes according to a sinusoidal function with a pre-defined period to capture the cyclical nature of demand and restocking~\cite{10.1007/978-3-642-99745-7_16}. 
%
The area is served by a 5G base station with omnidirectional antennas, transmitting at a fixed power of 5 dB to emulate average channel conditions and represent long-range indoor communication.
\begin{figure}
    \vspace{1mm}
    \centering
    \subfloat[\centering Corner]
    {{\includegraphics[clip, trim = 2cm 1cm 1.4cm 1.4cm, width=0.45\columnwidth]{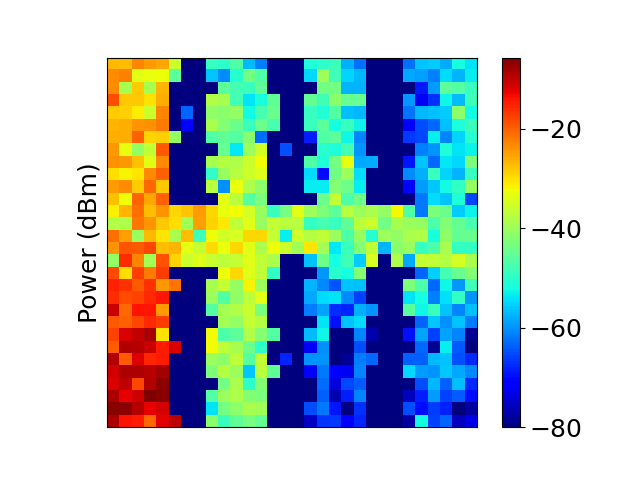}}}
    \subfloat[\centering Center]{{\includegraphics[clip, trim = 2cm 1cm 1.4cm 1.4cm, width=0.45\columnwidth]{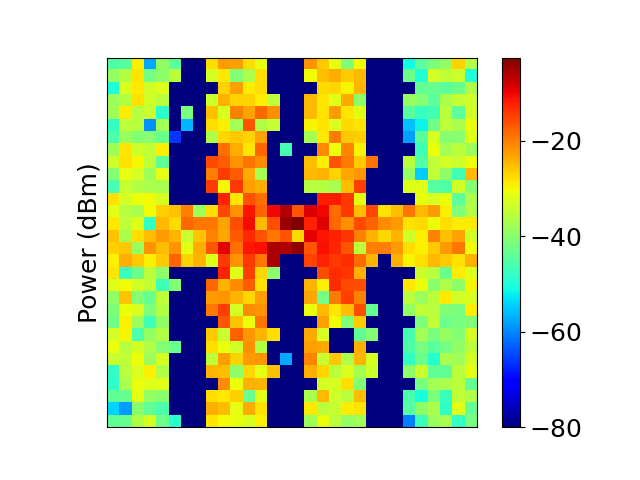}}}
    \qquad
    \vspace{1mm}
    \caption{\small Radio signal quality heatmaps in the warehouse with different antenna placements: (a) bottom-left corner, (b) center.}
    \label{fig:heatmap}
    \vspace{-1mm}
\end{figure}
\subsection{Architectural Assessment and Runtime Analysis}
\label{subsec:perf_arch}
In this subsection, we evaluate the performance of the proposed MapViT framework against several representative ML architectures to assess its effectiveness for dynamic RQMaps estimation. Specifically, we substitute the ViT module in each stage of MapViT with alternative architectures, namely, a MLP and a CNN, and compare their predictive accuracy, runtime efficiency, and generalization capability.
The experiments are conducted in a simulated industrial warehouse environment to analyze ($i$) the models’ ability to predict environmental changes (Stage 1) and radio quality maps (Stage 2), ($ii$) the trade-off between accuracy and computational cost, and ($iii$) the impact of scene dynamics on prediction performance. To ensure fairness, all architectures are configured to have approximately $0.6$ million trainable parameters.
The perceived channel quality in an indoor environment depends on both antenna placement and environmental settings. To validate this point, we simulate radio propagation within a warehouse using the \emph{Ray Tracer}, comparing two different antenna placements and a static distribution of goods. 
With reference to Fig.~\ref{fig:heatmap}(a), we can notice how the signal propagation is impaired by the physical objects located on the shelves when placing the transmission antenna in the bottom-left corner of the considered environment, leading to reduced signal strength, and therefore minimal coverage, on the opposite side of the room. Conversely, Fig.~\ref{fig:heatmap}(b) considers a scenario in which the base station antenna is centrally placed in the environment, leading to a symmetric signal distribution. While both antenna placements were explored, only the results for case Fig.~\ref{fig:heatmap}(a) will be shown in further analysis due to space constraints. 


\subsubsection{Stage 1 - Prediction of Environmental Changes}

To train MapViT for the prediction of depth maps (given the depth maps from previous time steps), we generate $10$k scenarios where each snapshot is linked with the previous one, i.e., the distribution of goods on the shelves and the corridors are evolving. 
The dataset is split according to a 9:1 ratio for training and testing sets. 
We adopt the Mean Squared Error (MSE) metric to train the models and choose the Adam optimizer with a learning rate $0.001$ to minimize the loss $\mathcal{L}$ over $50$ training epochs~\cite{KingBa15}. 
\begin{table}[ht]
\vspace{3mm}
\caption{\small Stage 1 prediction of environmental changes on PSNR.}
\centering
\begin{tabular}{|c|c|c|c|}
\hline
\textbf{Model} & ViT & CNN & MLP \\
\hline
\textbf{PSNR (dB)} & \textbf{29.01} & 27.53 & 24.06 \\
\hline
\end{tabular}
\label{tab:psnr_comparison}
\end{table}
We evaluate performance using the Peak Signal-to-Noise Ratio (PSNR) to quantify distortion in predicted maps, where higher PSNR values indicate better reconstruction quality~\cite{psnr}. 
As shown in Table~\ref{tab:psnr_comparison}, both ViT and CNN models achieve strong performance, with ViT slightly outperforming CNN. The MLP model yields substantially lower PSNR, indicating inability to capture spatial structure and temporal dynamics.
Both CNN and ViT effectively model variations,
with ViT showing roughly $1.5$ dB higher PSNR.

\subsubsection{Stage 2 - Prediction of RQMaps}
\begin{figure}[h]
    \centering
\includegraphics[width=0.85\columnwidth]{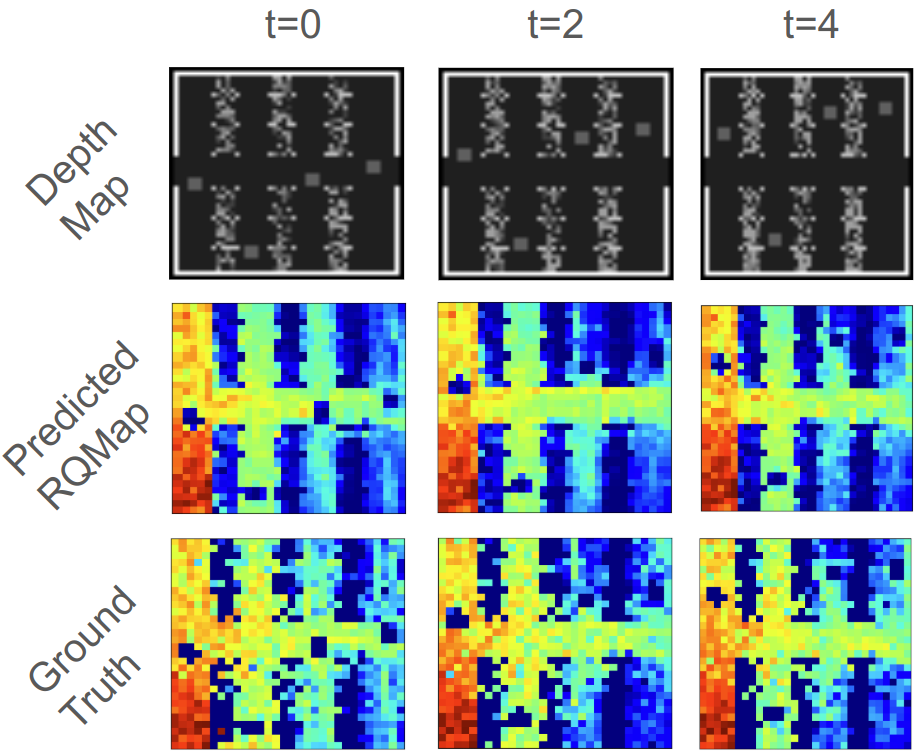}
    \caption{\small Evolution of depth maps, predicted RQMaps, and ground-truth RQMaps over time, produced by MapViT.}
    \label{fig:evolve}
    \vspace{5mm}
\end{figure}

The model is trained on $1$k different scenarios for the prediction of RQMaps (generated by ray tracer) given the corresponding depth maps.
The training configuration is the same as Stage 1 except for the number of epochs, which is extended to $100$.
Fig.~\ref{fig:evolve} illustrates the qualitative performance over a sequence of consecutive timestamps, demonstrating the MapViT's ability to adapt and predict in dynamic scenarios. The evolution of the depth maps (top row) and the predicted RQMaps (middle row) over time shows close alignment with the ground truth (bottom row), capturing the variability in the radio channel caused by moving objects in the scene.
This consistency enables accurate, real-time predictions and supports adaptive robot behavior in complex and rapidly changing environments with minimal computational overhead.

\begin{figure}[b]
    \centering
    \subfloat[\centering Inference time]{{\includegraphics[clip, trim = 0cm 0cm 0cm 0cm, width=0.5\columnwidth]{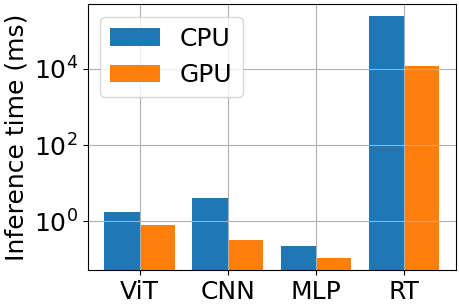}}}
     \qquad
    \subfloat[\centering Test regions]{{\includegraphics[trim = 0cm 0cm 0cm 0cm, width=0.35\columnwidth]{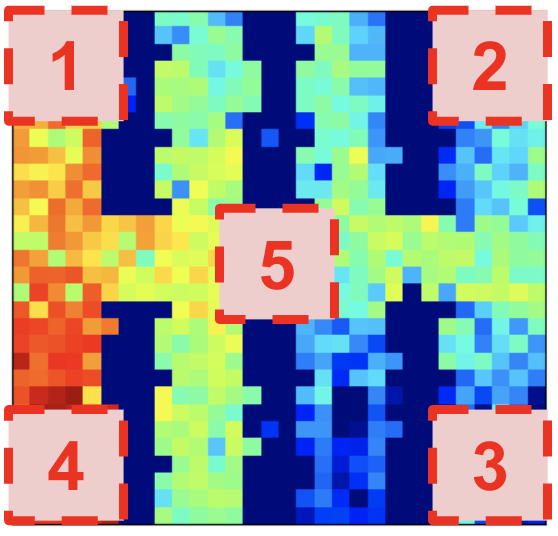}}}
    \qquad
    \vspace{2mm}
    \caption{\small (a) Compares Stage 2 inference time of ML-based approaches (ViT, CNN, MLP) against the ray tracer software (RT). (b) Depicts the evaluated regions in Table~\ref{table:regional_loss}.} 
    \label{fig:infer_region}
\end{figure}
Fig.~\ref{fig:infer_region}(a) compares the inference time and accuracy of the three models on both CPU and GPU platforms. The MLP achieves the fastest inference but sacrifices accuracy due to its limited spatial modeling capability. The CNN offers a balanced trade-off on GPUs due to parallelizable convolutional operations, but becomes the slowest on CPUs. The ViT obtains the highest accuracy while maintaining efficiency on CPUs, making it more suitable for resource-constrained devices. Overall, MLP favors speed, CNN excels on GPUs, and ViT provides the best balance between accuracy and efficiency across heterogeneous platforms.
We note that directly adopting the ray-tracing software for RQMap estimation comes with significant computational costs (e.g., up to $12$ seconds on an NVIDIA GeForce RTX 3080 GPU and approximately $4$ minutes on an Intel i7-12700K CPU under our test scenarios; see Fig.~\ref{fig:infer_region}(a)), hindering real-time decision-making process for both robots and networks. In contrast, the proposed MapViT framework achieves real-time prediction with substantially lower inference times ($\sim 1$ millisecond), making it well suited for real-time robotic applications and network optimization.
\
Beyond runtime and overall accuracy, we further analyze the spatial prediction quality across different regions in the warehouse. The evaluation based on regional PSNR across the predicted RQMaps in Table~\ref{table:regional_loss} reveals the superiority of the ViT model, which achieves the highest PSNR values not only in the global area but also in all five local areas labelled in Fig.~\ref{fig:infer_region}(b). This consistent advantage highlights the effectiveness of ViT in capturing the spatial distribution of radio signal strength, excelling especially in areas with strong signal presence (areas $1$ and $4$) and slightly surpassing CNN in regions with near-average signal strength (area $5$).
To assess the generalizability beyond the training distribution, we also evaluate their performance using out-of-distribution (OOD) data. In this test, shelves in areas $1$ and $4$ were removed completely to create new data reasonably distinct from the training data. The results, labeled as ViT* and CNN* in Table~\ref{table:regional_loss}, show that ViT retains a modest advantage over CNN.
%
\begin{table}[h]
\centering
\vspace{2mm}
\caption{\small Stage 2 performance comparison on PSNR (dB) for RQMaps prediction. See Fig.~\ref{fig:infer_region}(b) for the position of each area.}
\begin{tabular}[t]{|l|c|c|c|c|c|c|c|}
\hline
&Area 1&Area 2&Area 3&Area 4&Area 5&Global\\
\hline
ViT & \textbf{25.12} & \textbf{25.11} & \textbf{22.38} & \textbf{29.48} & \textbf{21.87} & \textbf{21.00}\\
CNN& 22.41 & 23.28 & 21.29 & 26.63 & 21.32 & 20.36 \\
MLP& 22.52 & 22.12 & 20.98 & 25.58 & 20.39 & 19.69 \\
\hline
ViT* & 21.70 & \textbf{19.24} & \textbf{17.19} & \textbf{24.68} & \textbf{15.93} & \textbf{14.50}\\
CNN* & 21.70 & 16.95 & 16.03 & 22.88 & 13.90 & 14.06\\
\hline
\end{tabular}
\label{table:regional_loss}
\vspace{-2mm}
\end{table}%

\subsection{Data Efficiency and Transferability}
\label{subsec:perf_data}
To investigate the data efficiency and generalization capability of the proposed framework for tasks besides the prediction of RQMaps, we explicitly evaluate the Stage 1 model of MapViT as a \textit{Geometry Foundation Model} (GFM).
Specifically, we compare the proposed two-stage scheme against a single-stage baseline trained directly in a supervised manner.
This comparison highlights the advantages of the decoupled learning process in terms of data efficiency.
In this experiment, MapViT is first trained on sequences of depth maps for $50$ epochs in a self-supervised fashion, as detailed in Sec.~\ref{sec:training}.
Subsequently, the model is fine-tuned for $50$ epochs using supervised learning on downstream spatial prediction tasks with labels involving geometry-related modalities, as summarized in Table~\ref{tab:propagation_maps_small}. Note that unlike pre-training, the fine-tuning stage can use independent samples rather than consecutive sequences, allowing greater flexibility in data usage.
\begin{figure}[h]
    \centering
\includegraphics[clip, trim = 0cm 0cm 0cm 0cm, width=1\columnwidth]{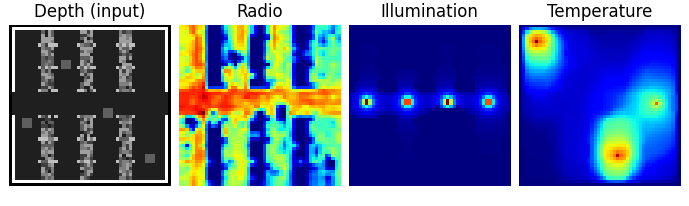}
    \caption{\small Examples of geometry-derived maps used in Table~\ref{tab:propagation_maps_small}. 
    }
    \label{fig:fields}
    \vspace{4mm}
\end{figure}
In these tasks, the model learns to translate depth maps into the corresponding geometry-derived maps, as exemplified in Fig.~\ref{fig:fields}.
Such task transferability is particularly relevant to the emerging 6G vision of multi-modal digital twins, where communication, sensing, and environment modeling are tightly integrated\cite{almadhor2025digital, GUERRA202480, 10198573}.
\begin{table}[b]
\centering
\caption{\small Three geometry-related propagation maps: Radio, illumination, and temperature map. Each map is derived from the same depth geometry but follows a different propagation model.}
\label{tab:propagation_maps_small}
\begin{tabular}{|l|c|c|c|}
\hline
\textbf{Property} & \textbf{Radio} & \textbf{Illum.} & \textbf{Temp.} \\
\hline
Physical model 
& Path loss
& Inverse-square
& Heat diffusion\\
\hline
Geometric relation
& Occlusion
& Distance 
& Connectivity\\
\hline
\end{tabular}
\vspace{-3mm}
\end{table}
The first task focuses on RQMap prediction, which is the same task as in Sec.~\ref{subsec:perf_arch}.
The second task involves predicting the illumination map, capturing light intensity distribution from four light sources positioned along the central corridor, while considering occlusions. 
The last task predicts the temperature map, which simulates steady-state heat diffusion across the same geometric layout from three heat sources (e.g., active engines). 
To evaluate the gain of geometry pre-training, models are initialized with and without the GFM weights and compared in terms of PSNR.
\begin{figure}[h]
    \vspace{-3mm}
    \centering
    \subfloat[\centering RQMap]{{\includegraphics[clip, trim = 0.4cm 0cm 0cm 0cm, width=0.5\columnwidth]{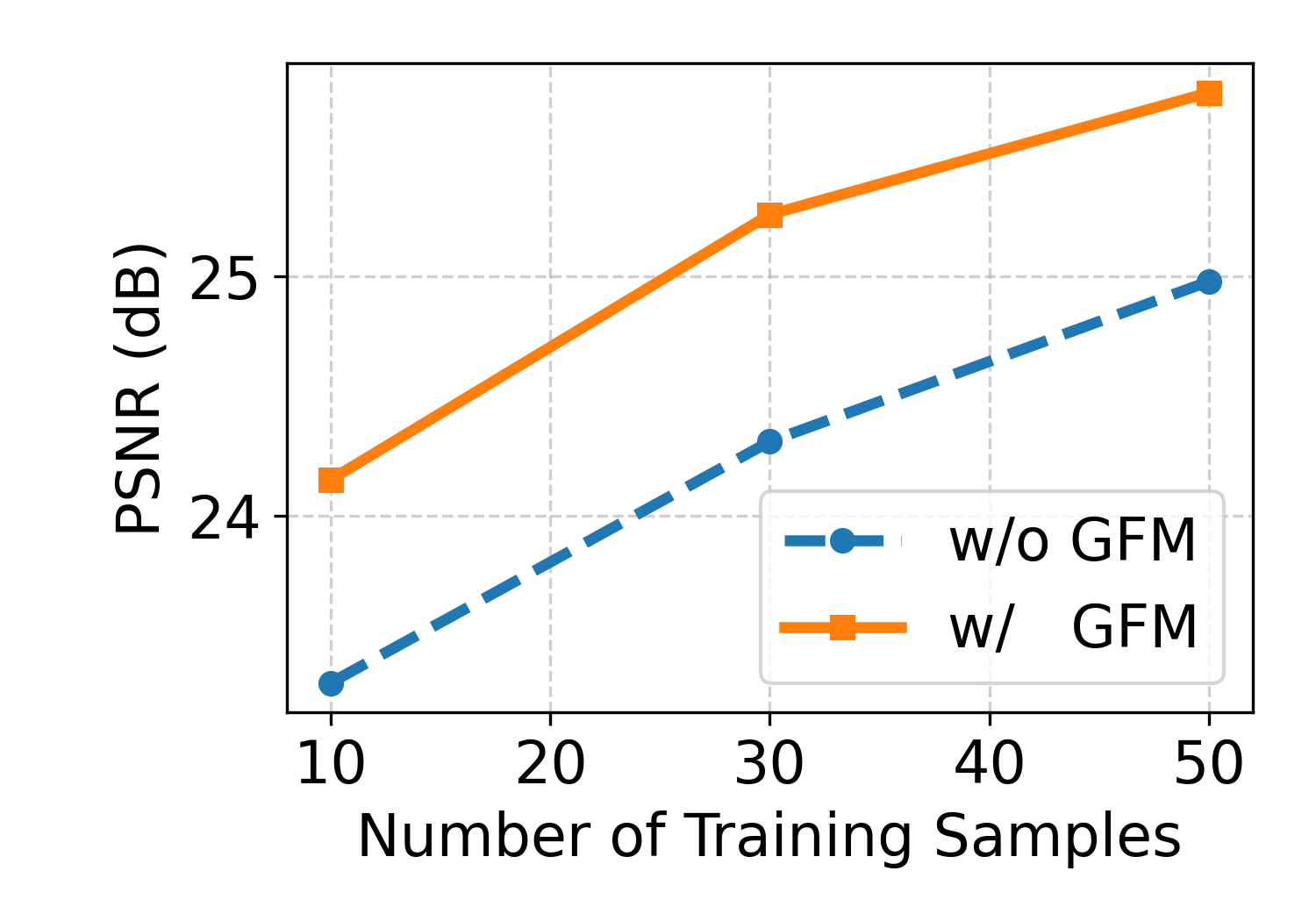}}}
    \subfloat[\centering Illumination map]{{\includegraphics[clip, trim = 0cm 0cm 0cm 0cm, width=0.51\columnwidth]{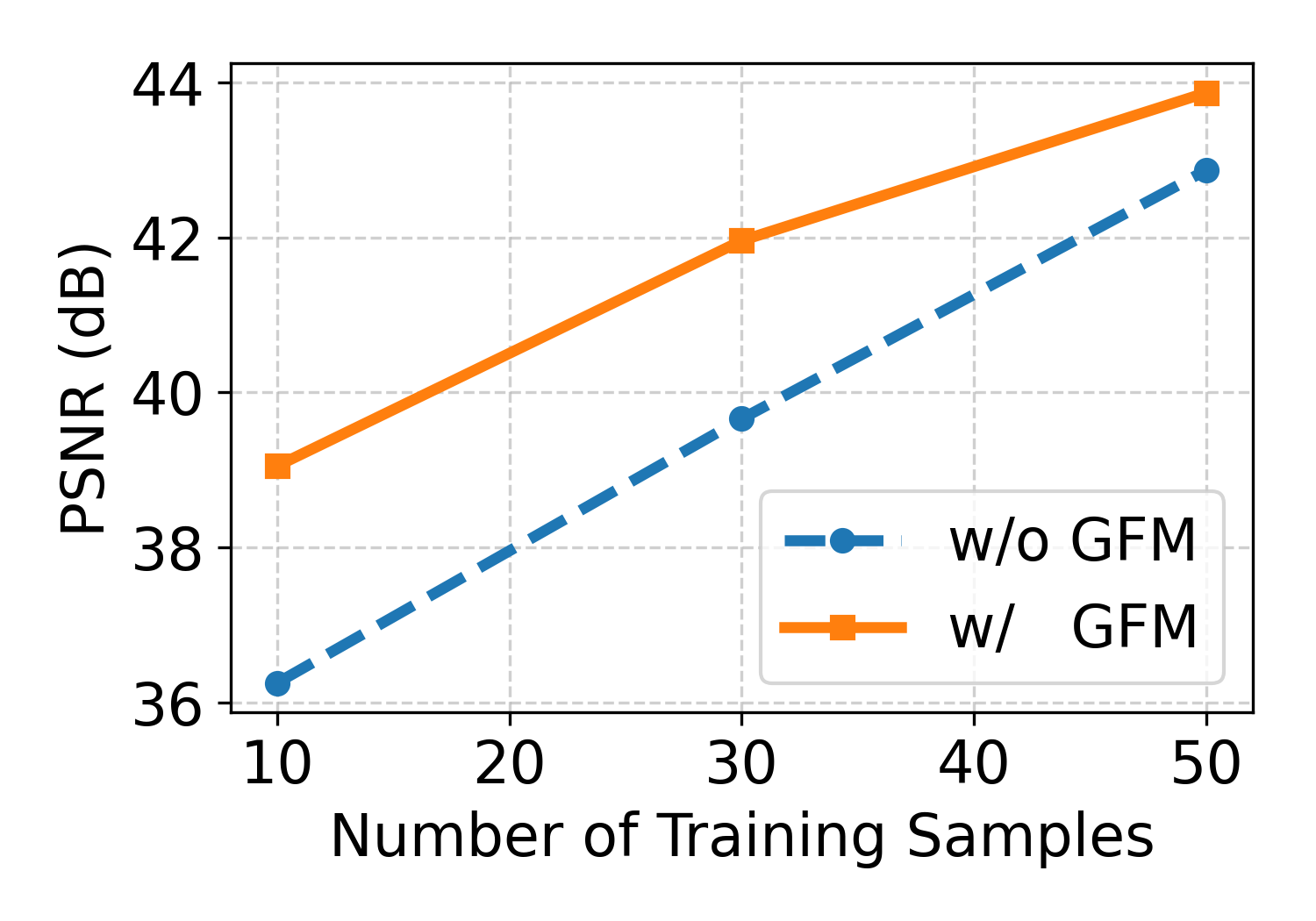}}}
    \\
    \vspace{4mm}
    \subfloat[\centering Temperature map]{{\includegraphics[clip, trim = 0cm 0cm 0cm 0cm, width=0.51\columnwidth]{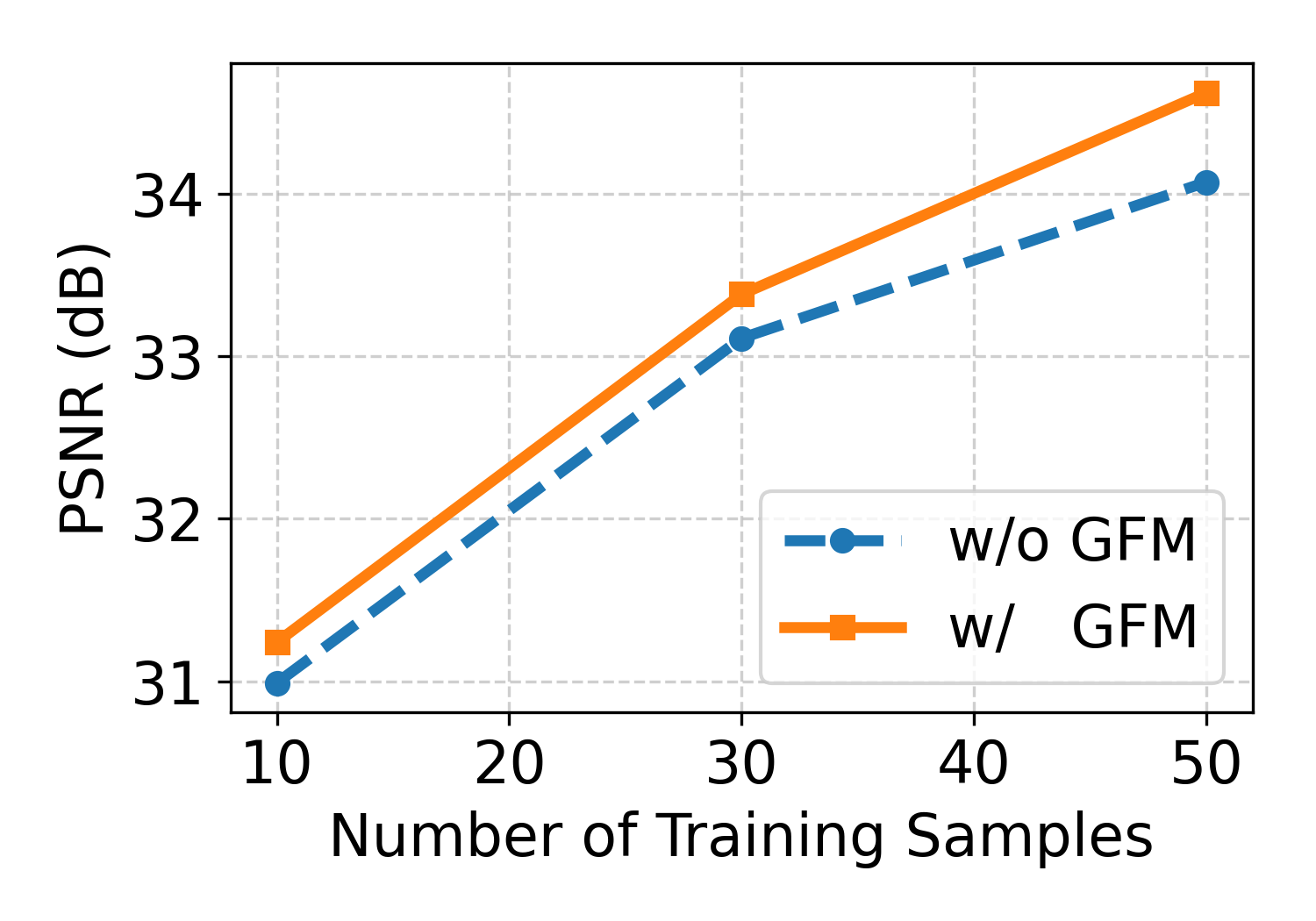}}}
    \subfloat[\centering Learning curves]{{\includegraphics[clip, trim = 0.5cm 0cm 0cm 0cm, width=0.51\columnwidth]{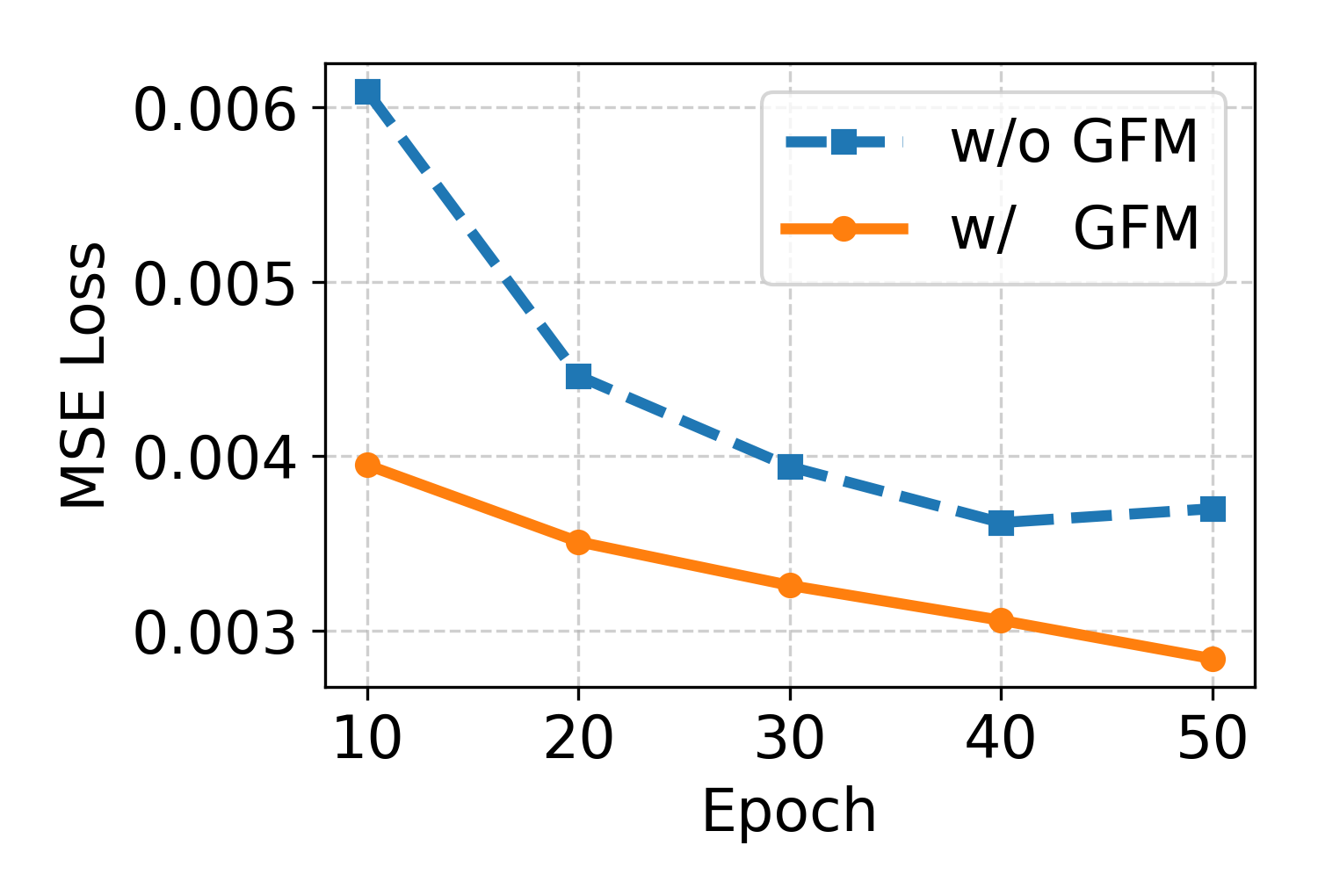}}}
    \vspace{3mm}
    \caption{\small Average PSNR and learning curves for various downstream tasks under limited labeled data. (a)–(c) PSNR obtained with and without GFM across different number of training samples. (d) Learning curves for RQMaps in terms of MSE using $50$ training samples.}
    \label{fig:fmpt}
    \vspace{3mm}
\end{figure}

Fig.~\ref{fig:fmpt} compares the PSNR performance and learning behaviors of models trained with and without the proposed geometry pre-training.
Across all downstream tasks, the pre-trained MapViT model consistently achieves higher PSNR values using few labeled samples.
This demonstrates that the GFM learned in Stage 1 provides a strong initialization that captures transferable spatial priors, enabling more data-efficient learning in downstream geometry-derived prediction tasks. The learning curves in Fig.~\ref{fig:fmpt}(d) further show that with GFM the learning converges faster with consistently lower testing loss, confirming the use of appropriate foundation models improves prediction accuracy and facilitates convergence.

\section{Conclusions and Future Work}
\label{sec:conclusion}
Radio map estimation presents a promising solution for enhancing the performance and adaptability of mobile robotic systems in dynamic environments. Through ML-based solutions, robots can effectively gain knowledge of dynamic environmental changes and adapt their navigation strategies and operation tasks to the changing wireless channel conditions.  To this end, we develop MapViT to estimate both the environmental changes and the radio quality variations, which achieves the best accuracy compared to ground truth while taking only $\sim\,1$ ms inference time. 
%
%
We evaluated the proposed framework using three representative architectures, including MLP, CNN, and ViT.
The results show that ViT achieves the best trade-off between accuracy and efficiency, making it suitable for deployment on resource-constrained platforms.
In addition, we show that the geometry foundation model learned from depth-map evolution significantly improves data efficiency, enabling higher accuracy and faster convergence in downstream geometry-derived prediction tasks with a limited amount of labeled data.
Future work will explore integrating MapViT into digital twins and applying advanced generative AI to unseen environments for improved generalizability.

\section*{Acknowledgment}
This research was partially funded by the Dutch 6G flagship project ``Future Network Services'',
and partially supported by the MultiX project from Smart Networks and Services Joint Undertaking (SNS JU) under the European Union’s Horizon Europe research and innovation programme (Grant 101192521).


\bibliographystyle{IEEEtran}
\bibliography{biblio}
\end{document}